\pdfoutput=1
\documentclass[fleqn]{2017SCGE}
\usepackage[linkcolor=blue,citecolor=blue]{hyperref}

\begin{document}

	\ensubject{subject}

	\ArticleType{Article}
	\Year{2020}
	\Month{February}
	\Vol{xx}
	\No{xx}
	\DOI{https://doi.org/10.1007/s11431-020-1706-9}
	\ArtNo{xxx}
	
	\title{Mapping the magnetic field in the solar corona through magnetoseismology}

	\author[1]{YANG ZiHao}{}%
	\author[1,2]{TIAN Hui}{huitian@pku.edu.cn}
	\author[3]{TOMCZYK Steven}{}
	\author[4]{MORTON Richard}{}
	\author[2]{\\BAI XianYong}{}
	\author[5,6]{SAMANTA Tanmoy}{}
	\author[1,7]{CHEN YaJie}{}

	\AuthorMark{Yang Z}
	
	\AuthorCitation{Yang Z, Tian H, Tomczyk S, et al.}
	
	\address[1]{School of Earth and Space Sciences, Peking University, Beijing 100871, China; huitian@pku.edu.cn}
	\address[2]{Key Laboratory of Solar Activity, National Astronomical Observatories, Chinese Academy of Sciences, Beijing 100012, China}
	\address[3]{High Altitude Observatory, National Center for Atmospheric Research, Boulder, CO 80307, USA}
	\address[4]{Department of Mathematics, Physics and Electrical Engineering, Northumbria University, Newcastle Upon Tyne NE1 8ST, UK}
	\address[5]{Department of Physics and Astronomy, George Mason University, Fairfax, VA 22030, USA}
	\address[6]{Johns Hopkins University Applied Physics Laboratory, Laurel, MD 20723, USA}
	\address[7]{Max Planck Institute for Solar System Research, 37077 G\"{o}ttingen, Germany}

	
	\abstract{Magnetoseismology, a technique of magnetic field diagnostics based on observations of magnetohydrodynamic (MHD) waves, has been widely used to estimate the field strengths of oscillating structures in the solar corona. 
	However, previously magnetoseismology was mostly applied to occasionally occurring oscillation events, providing an estimate of only the average field strength or one-dimensional distribution of field strength along an oscillating structure. 
	This restriction could be eliminated if we apply magnetoseismology to the pervasive propagating transverse MHD waves discovered with the Coronal Multi-channel Polarimeter (CoMP). 
	Using several CoMP observations of the Fe \sc{xiii}\rm{} 1074.7 nm and 1079.8 nm spectral lines, we obtained maps of the plasma density and wave phase speed in the corona, 
	which allow us to map both the strength and direction of the coronal magnetic field in the plane of sky. We also examined distributions of the electron density and magnetic field strength, 
	and compared their variations with height in the quiet Sun and active regions. Such measurements could provide critical information to advance our understanding of the Sun's magnetism and the magnetic coupling of the whole solar atmosphere.}

	\keywords{Solar corona, Solar magnetic field, Waves, Magnetoseismology}

	\maketitle

	\begin{multicols}{2}
		\section{Introduction}\label{section1}
		
\begin{figure}[H]
			\centering
			\includegraphics[scale=0.8]{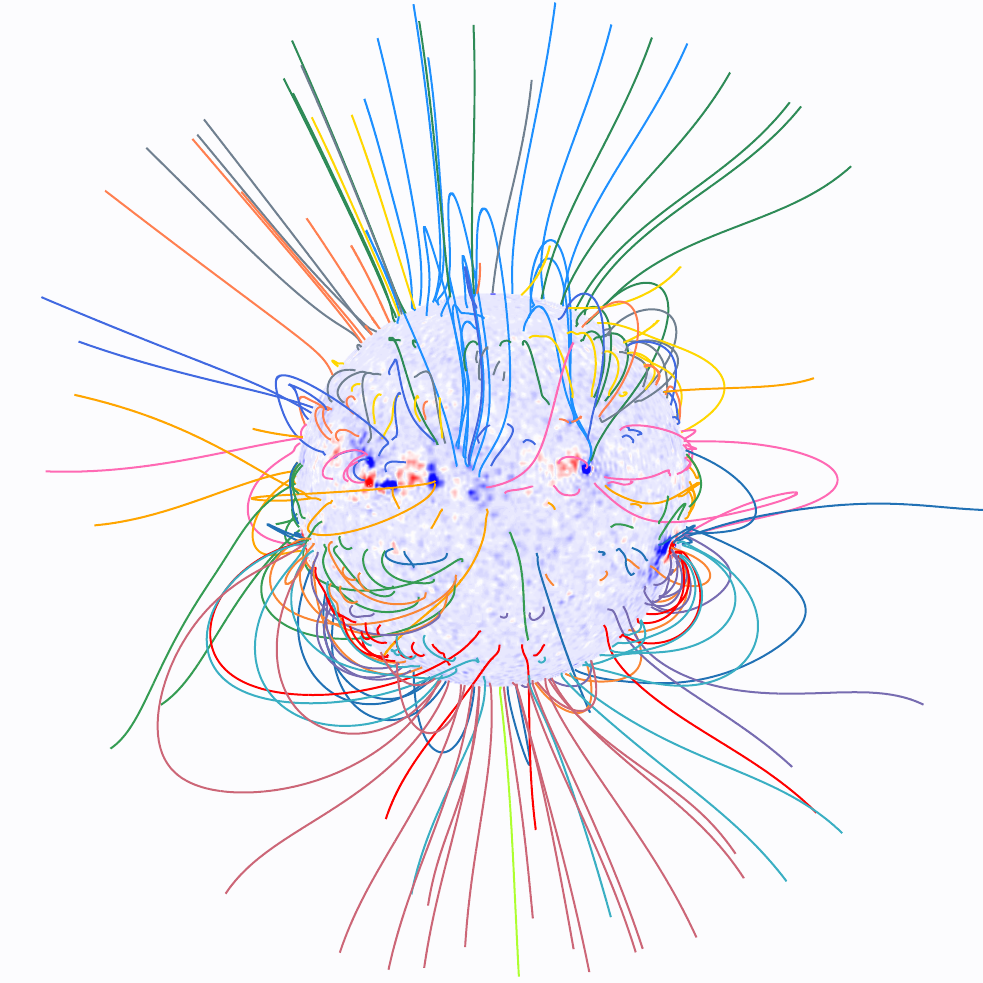}
			\caption{Global coronal magnetic field on Oct 14, 2016 obtained from the PFSS model. The colored lines represent selected magnetic field lines. The synoptic photospheric magnetogram used for the PFSS model is shown in the middle, and the red and blue patches represent different polarities of the longitudinal magnetic field.} 
			\label{fig:fig1}
		\end{figure}

		Originating from the solar interior, the solar magnetic field extends to the solar surface and couples different layers of the solar atmosphere (\cref{fig:fig1}). 
		Because of this, information on the magnetic field of the whole atmosphere is required to study the interplay between the solar plasma and magnetic field. 
		However, routine and reliable measurements of the solar magnetic field have only been achieved at the photospheric level (e.g., \cite{ref66,ref67}). 
		We still do not have a precise knowledge of the magnetic field in the upper solar atmosphere, especially the corona, which impedes our complete understanding of a wide range of phenomena including the solar cycle, solar eruptions and heating of the coronal plasma.
		
		Without routine measurements of the coronal magnetic field, extrapolations from the observed photospheric magnetograms often serve as an important tool for reconstruction of coronal magnetic field structures (e.g., \cite{ref1,ref2,ref3,ref4,ref68}). Extrapolation models can be classified into potential field, linear force-free field and non-linear force-free field extrapolations. \cref{fig:fig1} shows an example of the global coronal magnetic field structures obtained from the most frequently used potential field source surface (PFSS) model. In addition, magnetohydrostatic (e.g., \cite{ref5,ref6}) and magnetohydrodynamic (MHD) models (e.g., \cite{ref7,ref8}) have also been developed to reconstruct three-dimensional (3D) magnetic field structures in the solar atmosphere. Although widely used for investigations of the coronal magnetism, these models are highly dependent on various assumptions, which are not always valid on the Sun.

        Direct measurements of the coronal magnetic field through spectropolarimetric observations have been attempted in the past two decades. The linear polarization of some coronal forbidden lines is sensitive to the magnetic field orientation in the plane of sky (POS) (e.g., \cite{ref9,ref10,ref11}), and the degree of linear polarization could be used to diagnose some magnetic structures such as flux ropes \cite{ref12,ref13,ref14}. Due to the weak field in the corona, circular polarization signals associated with the longitudinal Zeeman effect are usually very weak. In order to measure the line-of-sight (LOS) component of the coronal magnetic field, often a long integration time is required to increase the signal-to-noise ratio (S/N) of the circular polarization data. Only a couple of successful measurements have been performed using this method \cite{ref15,ref16}.
        
        Besides spectropolarimetric measurements, several other techniques have also been developed to probe the coronal magnetic field. For instance, the standoff distance method could be used to infer the field strengths along the paths of shocks driven by fast coronal mass ejections (CMEs) \cite{ref17}. Radio spectral observations of type III bursts, spike structures associated with type IV bursts and zebra patterns could also provide estimates of the magnetic field in the coronal source regions of the radio emission \cite{ref18,ref19,ref20}. In addition, when the emission mechanisms are known, microwave imaging at one or more frequencies could be used to produce maps of the coronal magnetic field strength in limited regions \cite{ref21,ref22,ref23,ref24,ref25,ref26}. More recently, the phenomenon of magnetic-field-induced transition (MIT) \cite{ref27} has caught the attention of the solar physics community, and theoretical and laboratory investigations have demonstrated the potential of MIT lines in measurements of the coronal magnetic field \cite{ref28,ref29,ref70}.
        
        Another frequently used approach for measurements of the coronal magnetic field is coronal seismology or magnetoseismology, which refers to magnetic field diagnostics based on observations of MHD waves or oscillations (e.g., \cite{ref30}). For transverse kink oscillations, the magnetic field strengths in or outside oscillating structures could be derived if the local density is assumed or estimated \cite{ref31,ref32,ref33,ref34,ref35,ref36,ref37,ref38}. However, these rare single-oscillation events are often related to flares or CMEs, and their observations can provide an estimate of only the average field strength or one-dimensional (1D) distribution of field strength along an oscillating structure. This restriction could be eliminated if we apply magnetoseismology to more ubiquitous and continuous oscillations or waves in the corona. At least two types of such oscillations/waves are known to exist in the corona: the decayless/persistent standing transverse waves in coronal loops observed through extreme-ultraviolet (EUV) imaging and spectroscopic observations \cite{ref39,ref40,ref41,ref42}, and the pervasive propagating transverse waves observed with the Coronal Multi-channel Polarimeter (CoMP, \cite{ref43}) \cite{ref44,ref45,ref46,ref47,ref48,ref49}. These ubiquitous oscillations/waves, especially the latter, are potentially important for continuous diagnostics of coronal magnetic field \cite{ref47,ref50}. By applying the technique of magnetoseismology to the pervasive propagating waves observed with CoMP, we could map the magnetic field in the corona. The first attempt was made for a trans-equatorial loop system, though the magnetic field magnitude was obtained at only a limited number of pixels in the bottom part of the loop system \cite{ref51}. Recently, using CoMP observations on October 14, 2016, we managed to perform the first measurement of the global coronal magnetic field in the POS \cite{ref52}.
        
        In this paper, we will present analysis results for more datasets using the same technique. We will also present maps of the POS magnetic field orientation measured from wave observations and the magnetic azimuth derived from linear polarization observations. In addition, we will examine distributions of the electron density and magnetic field strength, and compare their variations with height in the quiet Sun and active regions.

		\section{Instruments and Data}\label{sec:2}
		
\begin{figure}
			\centering
			\includegraphics[scale=0.8]{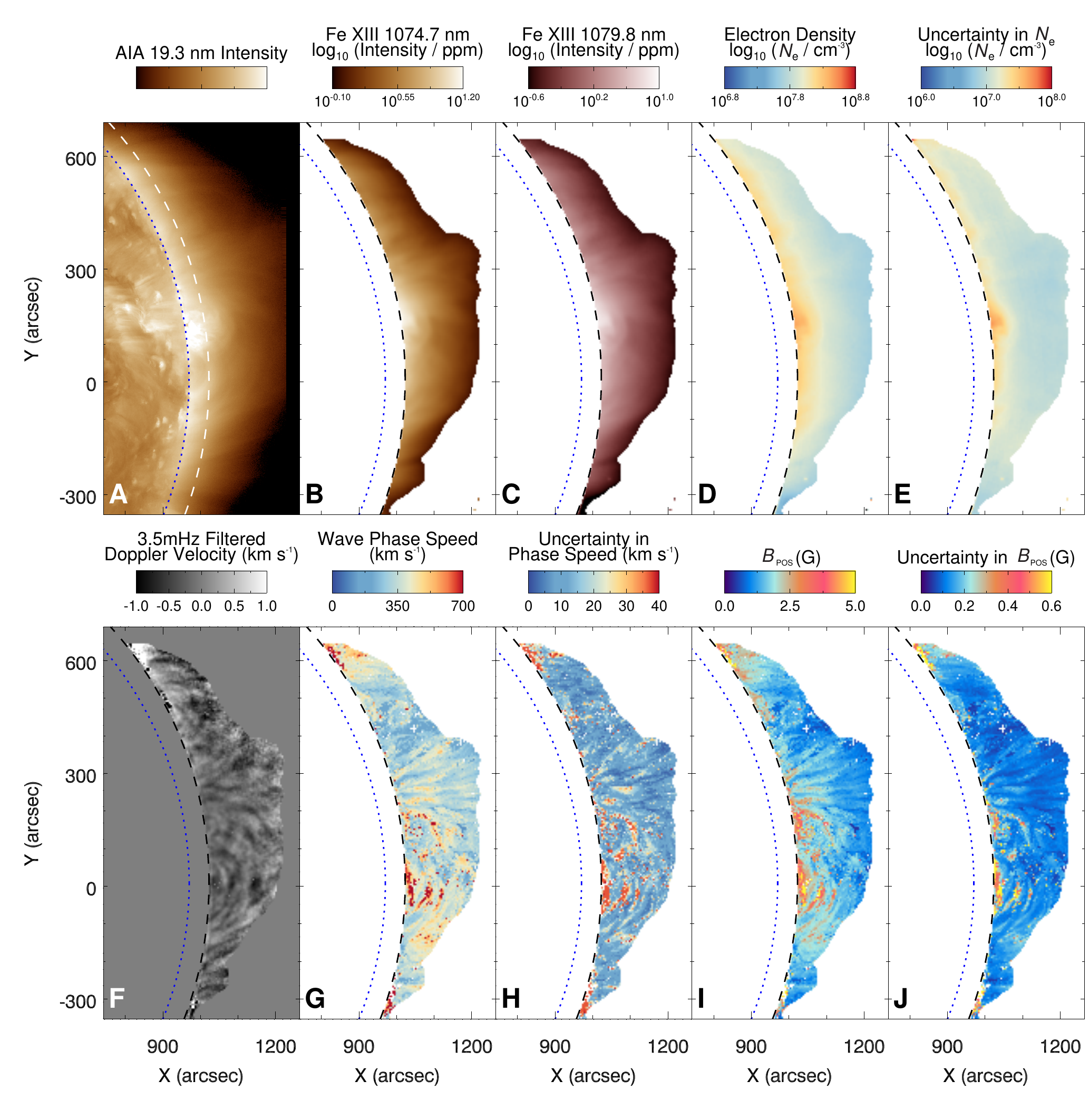}
			\caption{Analysis results for the observation on November 3, 2016 (dataset D2). The dotted and dashed curves mark the solar limb and the inner boundary of the CoMP FOV, respectively. (A) The AIA 19.3 nm image taken at 20:00:07 UT. (B)(C): The intensity maps of Fe \sc{xiii}\rm{} 1074.7 nm and 1079.8 nm. (D)-(E): The maps of electron number density and its uncertainty. (F): The filtered Doppler velocity map at 20:34:54 UT. (G)-(H): The maps of phase speed and its uncertainty. (I)-(J): The maps of derived $B_\text{POS}$ and its associated uncertainty.} 
			\label{fig:fig2}
		\end{figure}

\begin{figure}
			\centering
			\includegraphics[scale=0.8]{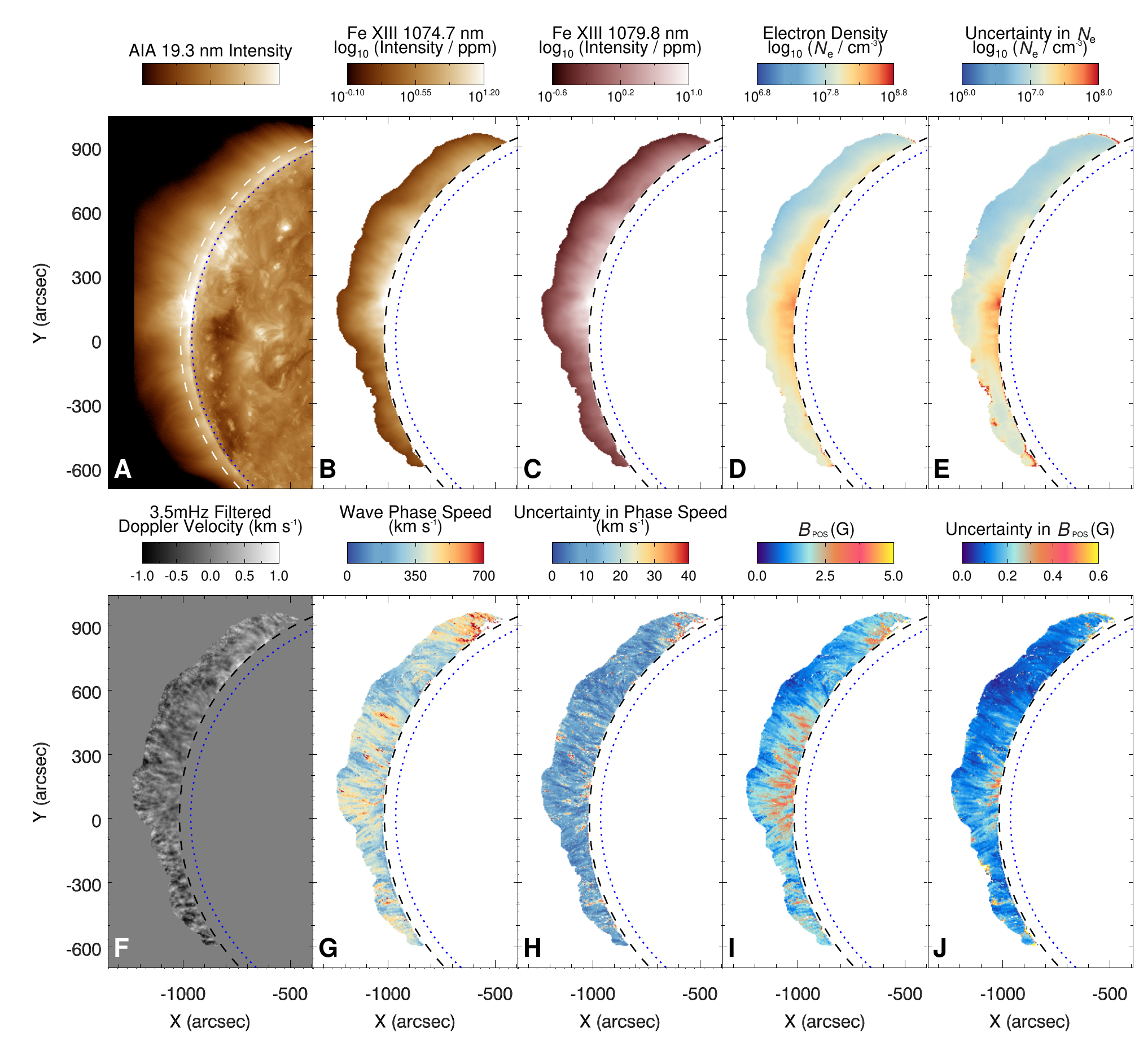}
			\caption{Same as \cref{fig:fig2}, but for the observation on March 20, 2017 (dataset D3).} 
			\label{fig:fig3}
		\end{figure}

		The CoMP is a 20-cm aperture coronagraph mounted at Mauna Loa Solar Observatory in Hawaii. It can perform spectropolarimetric observations at infrared wavelengths. We used the data sampled at several wavelength positions across the spectral profiles of Fe \sc{xiii}\rm{} 1074.7 nm and 1079.8 nm on October 14, 2016 (dataset D1), November 3, 2016 (dataset D2) and March 20, 2017 (dataset D3). These three observations were taken under relatively good and stable observing conditions. The spatial sampling is $\sim$4.35$^{\prime\prime}$. The field-of-view (FOV) is about 1.05-1.35 solar radii from the solar center. We used the Stokes-I profiles of the two lines for density diagnostics, the Stokes-I data of Fe \sc{xiii}\rm{} 1074.7 nm for wave tracking, and the Stokes Q and U data of Fe \sc{xiii}\rm{} 1074.7 nm for calculation of magnetic azimuth. The Doppler velocity and peak intensity were obtained through an analytical Gaussian fitting to the three-point intensity (Stokes-I) profile at each spatial pixel \cite{ref53}. The details of the three observations are summarized in \cref{tab:tab1}. For dataset D1, intensity images of the two Fe \sc{xiii}\rm{} lines have been presented in our previous work \cite{ref52}. \cref{fig:fig2}(B)-(C) and \cref{fig:fig3}(B)-(C) show the Fe \sc{xiii}\rm{} intensities for the other two datasets. Note that for datasets D2 and D3, we present results derived from only the good-quality data in parts of the global corona. We only show pixels where the Fe \sc{xiii}\rm{} 1074.7 nm peak intensity is higher than 1.0 ppm (millionth of the solar disk intensity). In addition, we excluded pixels where the phase speed (see below) exceeds $2000\ \text{km}\ \text{s}^{-1}$ from the maps of phase speed and magnetic field strength.

		For comparison, we also used the simultaneously obtained 19.3 nm images observed by the Atmospheric Imaging Assembly (AIA, \cite{ref54}) on the Solar Dynamics Observatory (SDO). For dataset D1, again we refer the chosen AIA image to our previous work \cite{ref52}. For dataset D2, we chose the AIA 19.3 nm image observed at 20:00:07 UT on November 3, 2016 (\cref{fig:fig2}(A)). For dataset D3, the AIA 19.3 nm image obtained at 19:00:18 UT on March 20, 2017 was used (\cref{fig:fig3}(A)). These images have a spatial sampling of $\sim$0.6$^{\prime\prime}$.

 \end{multicols}	
 
 \begin{table}[H]
 \centering
 \caption{Details of the CoMP observations used in our study.} 
\begin{tabular}{|l|l|l|l|l|}
\hline
Datasets              & \begin{tabular}[c]{@{}l@{}}Time range for \\ density diagnostics\end{tabular} & \begin{tabular}[c]{@{}l@{}}Frame number\\ for density \\ diagnostics\end{tabular} & \begin{tabular}[c]{@{}l@{}}Time range for\\ wave tracking/ \\azimuth calculation\end{tabular} & \begin{tabular}[c]{@{}l@{}}Frame number for\\ wave tracking/ \\azimuth calculation\end{tabular} \\ \hline
D1 (October 14, 2016) & 19:24 UT - 20:17 UT                                                           & \begin{tabular}[c]{@{}l@{}}1074.7 nm: 49\\ 1079.8 nm: 10\end{tabular}          & 20:39 UT - 21:26 UT                                                            & 1074.7 nm: 94                                                                               \\ \hline
D2 (November 3, 2016) & 19:14 UT - 20:20 UT                                                           & \begin{tabular}[c]{@{}l@{}}1074.7 nm: 63\\ 1079.8 nm: 12\end{tabular}          & 20:34 UT - 21:37 UT                                                            & 1074.7 nm: 125                                                                              \\ \hline
D3 (March 20, 2017)   & 18:27 UT - 19:43 UT                                                           & \begin{tabular}[c]{@{}l@{}}1074.7 nm: 78\\ 1079.8 nm: 14\end{tabular}          & 19:53 UT - 20:44 UT                                                            & 1074.7 nm: 101                                                                              \\ \hline
\end{tabular}
\label{tab:tab1}
\end{table}

		 \noindent
		
\begin{multicols}{2}		
		\section{Analysis results}\label{sec:3}
		\subsection{Coronal magnetic field measurements through magnetoseismology}
		Previous CoMP observations have shown that the pervasive propagating transverse waves, manifested as propagating perturbations in the Fe  \sc{xiii}\rm{} 1074.7 nm Doppler velocity sequences, often have a power spectrum peaked around 3.5 mHz (corresponding to a period of $\sim$5 min) \cite{ref44,ref45,ref47,ref49}. These transverse waves are identified as kink or Alfv\'{e}nic waves. Considering the lower-beta coronal environment and the moderate spatial resolution of CoMP ($\sim$9$^{\prime\prime}$), an appropriate expression of the wave phase speed ($c_k$) is \cite{ref47,ref50,ref51,ref52}
		
		\begin{equation}
		c_k=\frac{B}{\sqrt{\mu_0\left<\rho\right>}}
		\label{Eq:1}
		\end{equation}
		where $B$, $\mu_0$ and $\left<\rho\right>$ are the magnetic field strength, magnetic permeability of a vacuum and plasma density averaged within a spatial pixel, respectively. This equation shows that the field strength can be directly calculated once the phase speed and density are known.
		
		To calculate the phase speed and plasma density, we utilized the wave tracking procedure that was developed around 2007 and further improved in the past decade \cite{ref44,ref45,ref47,ref52,ref55}, and the same line ratio method for density diagnostic in our previous work \cite{ref52}. Detailed descriptions of these methods and estimations of the associated uncertainties can be found in the publications mentioned above. Here we just briefly describe the diagnostic procedures and present the results. 
		
		\begin{figure}[H]
			\centering
			\includegraphics[scale=0.8]{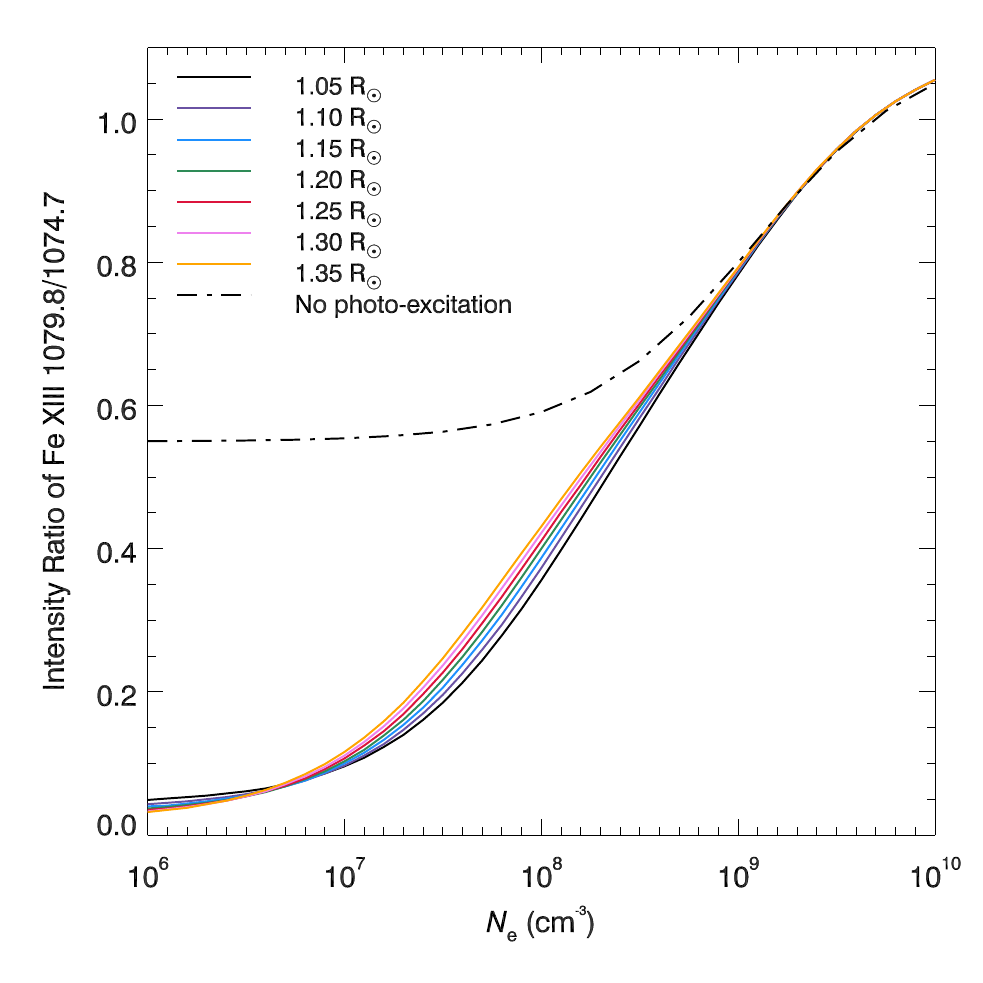}
			\caption{The theoretical relationship between line ratio and electron density. The dot-dashed curve is the result without consideration of photo-excitation. The colored solid curves show results at different heights when taking into account both collisional excitation and photo-excitation.} 
			\label{fig:fig4}
		\end{figure} 
		
		The intensity ratio of the Fe \sc{xiii}\rm{} 1079.8 nm / 1074.7 nm line pair is known to be sensitive to electron density ($N_\text{e}$). We utilized the CHIANTI database version 9.0 \cite{ref56,ref57} to generate the theoretical relationship between the Fe \sc{xiii}\rm{} 1079.8/1074.7 line ratio and electron density in the range of 1.05 $\text{R}_\odot$ to 1.35 $\text{R}_\odot$, considering both collisional excitation and photo-excitation. To include photo-excitation, a uniform spherical blackbody radiation source with a typical solar surface temperature is assumed \cite{ref58}. \cref{fig:fig4} shows the theoretical curves at several heights. As a comparison, the theoretical curve without consideration of photo-excitation is also plotted. From these theoretical curves, we obtained maps of the electron number density and the associated uncertainty. The results are presented in our previous work \cite{ref52} (dataset D1) and in \cref{fig:fig2}(D)-(E) (dataset D2) and \cref{fig:fig3}(D)-(E) (dataset D3). Results from all datasets show that the electron number density mostly falls in the range of $10^{7.5}-10^{8.5}\ \text{cm}^{-3}$. Considering the typical coronal composition, the average plasma mass density ($\left<\rho\right>$) can be calculated as $1.2N_\text{e} m_\text{p}$, where $m_\text{p}$ is the proton mass (e.g., \cite{ref40}).

		
		The phase speed is obtained through the wave tracking procedure. For each spatial pixel, we first interpolated the Doppler velocity time series to achieve a 30-s regular cadence, then filtered the interpolated time series to extract the 3.5-mHz (1.5-mHz FWHM) component. After that, a $41\times 41$ pixel box was assigned around each spatial pixel, and the coherence between the time series at this and surrounding pixels was calculated. The high coherence region is usually elongated, and the orientation of its elongation was defined as the wave propagation direction/angle. After performing this calculation for all pixels, we can obtain a map of wave propagation angle. For the purpose of comparison, we also calculated the magnetic azimuth from linear polarization observations using the following equation
		
		\begin{equation}
		\phi=\frac{1}{2}\tan^{-1}(\frac{U}{Q})
		\label{Eq:2}
		\end{equation}
		
\begin{figure}[H]
			\centering
			\includegraphics[scale=0.8]{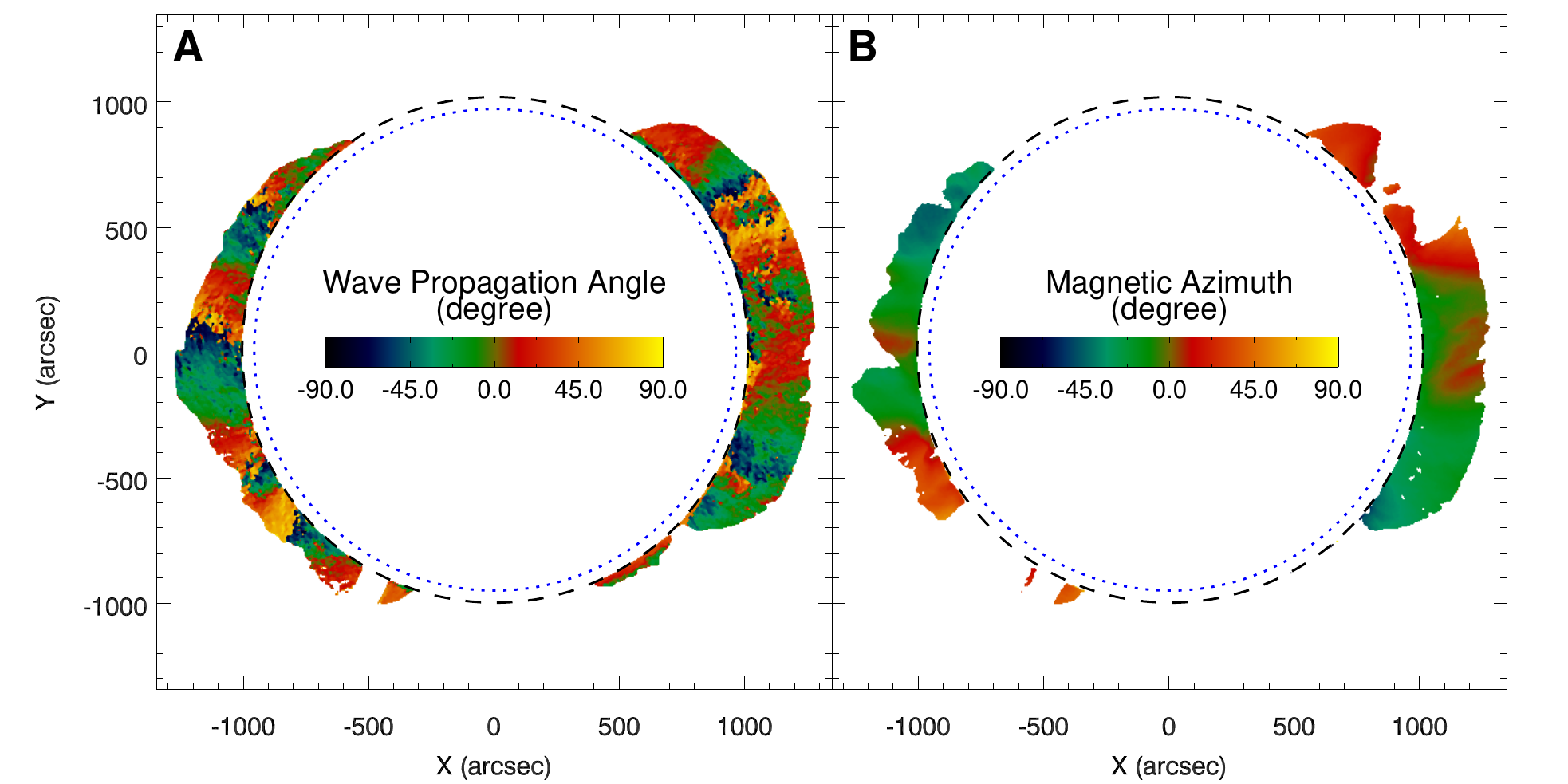}
			\caption{The maps of wave propagation angle (A) and magnetic azimuth (B) for the observation on October 14, 2016 (dataset D1). The dotted and dashed curves mark the solar limb and the inner boundary of the CoMP FOV, respectively. Pixels with a linear polarization degree lower than 0.06 have been removed from the azimuth map.} 
			\label{fig:fig5}
		\end{figure}

\begin{figure}[H]
			\centering
			\includegraphics[scale=0.8]{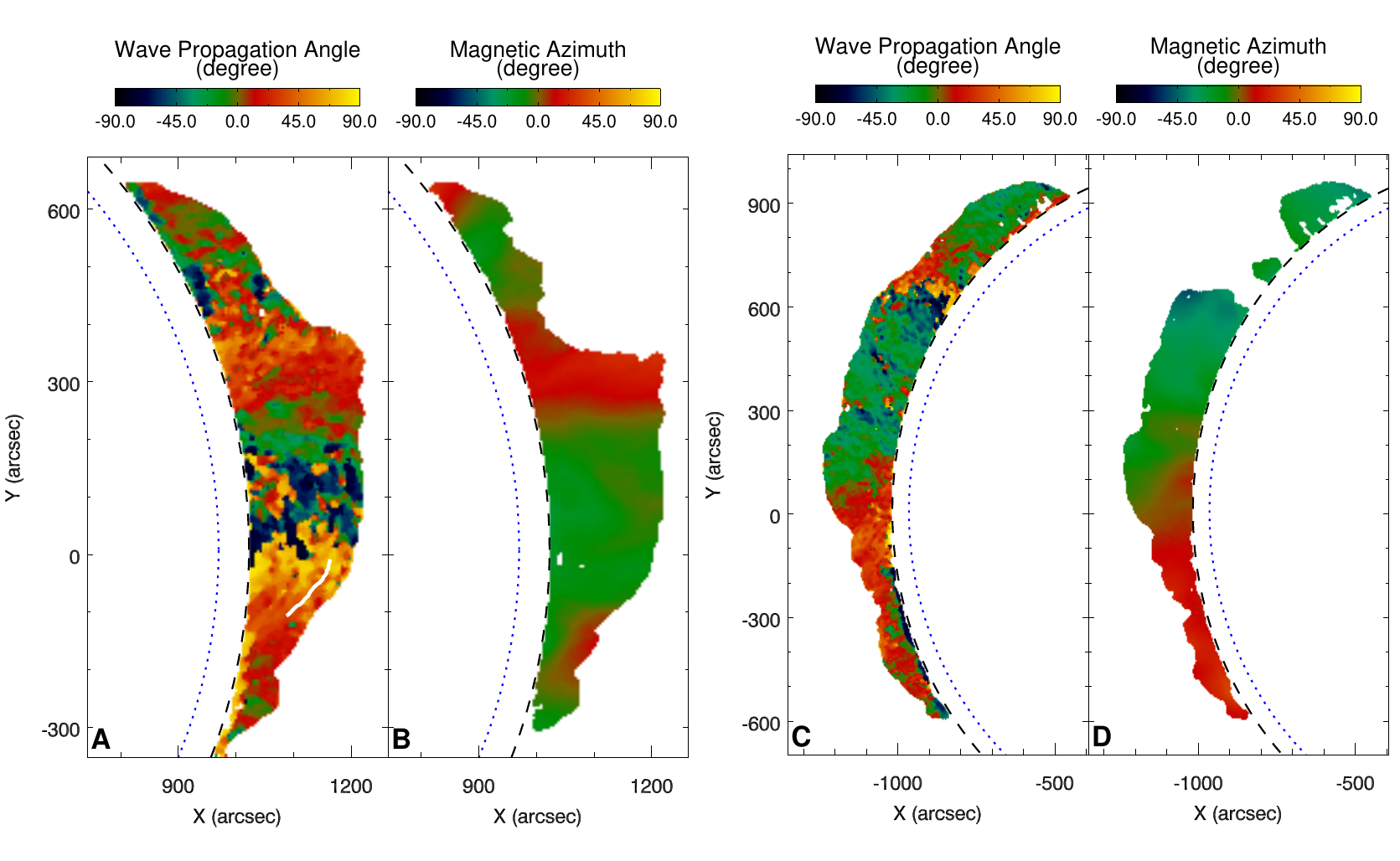}
			\caption{Same as \cref{fig:fig5}, but for datasets D2 (A-B) and D3 (C-D). The white curve in panel A shows a sample wave path used for the calculation of phase speed in \cref{fig:fig7}.} 
			\label{fig:fig6}
		\end{figure}

		The linear polarization measurements have an intrinsic 180$^{\circ}$ ambiguity. \cref{fig:fig5} shows the maps of wave propagation angle and magnetic azimuth for dataset D1. Similar maps for datasets D2 and D3 are depicted in \cref{fig:fig6}. Both the wave propagation direction and magnetic azimuth were measured relative to the east-west direction and are restricted within the range of -90$^{\circ}$ and 90$^{\circ}$ due to the 180$^{\circ}$ ambiguity. In the maps of magnetic azimuth, we only show pixels with a linear polarization degree greater than 0.06, because a lower linear polarization degree is subject to a greater uncertainty in the calculated azimuth \cite{ref59}. The patterns of wave propagation angle and magnetic azimuth are generally similar for all the three datasets, suggesting that both parameters could provide information on the POS direction of the magnetic field. However, distinct differences are seen at several locations, especially locations where apparent loop structures are present, e.g., y positions of -100$^{\prime\prime}$ to 200$^{\prime\prime}$ in \cref{fig:fig6}(A). In these regions, the wave propagation directions are generally consistent with the loop structures visible in the corresponding AIA intensity images. Note that the abrupt changes of wave propagation angles at some pixels in these regions are actually caused by the 180$^{\circ}$ ambiguity of wave angle, e.g., wave angles with values close to -90$^{\circ}$ and 90$^{\circ}$ actually represent similar wave propagation directions. In some parts of these regions, the angle between the local magnetic field and the solar radial direction could exceed the Van Vleck angle (54.74$^{\circ}$, \cite{ref60}), and thus the measured azimuth is subject to the 90$^{\circ}$ Van Vleck ambiguity. This effect appears to be responsible for some differences between the calculated wave angle and magnetic azimuth.
		
\begin{figure}[H]
			\centering
			\includegraphics[scale=0.7]{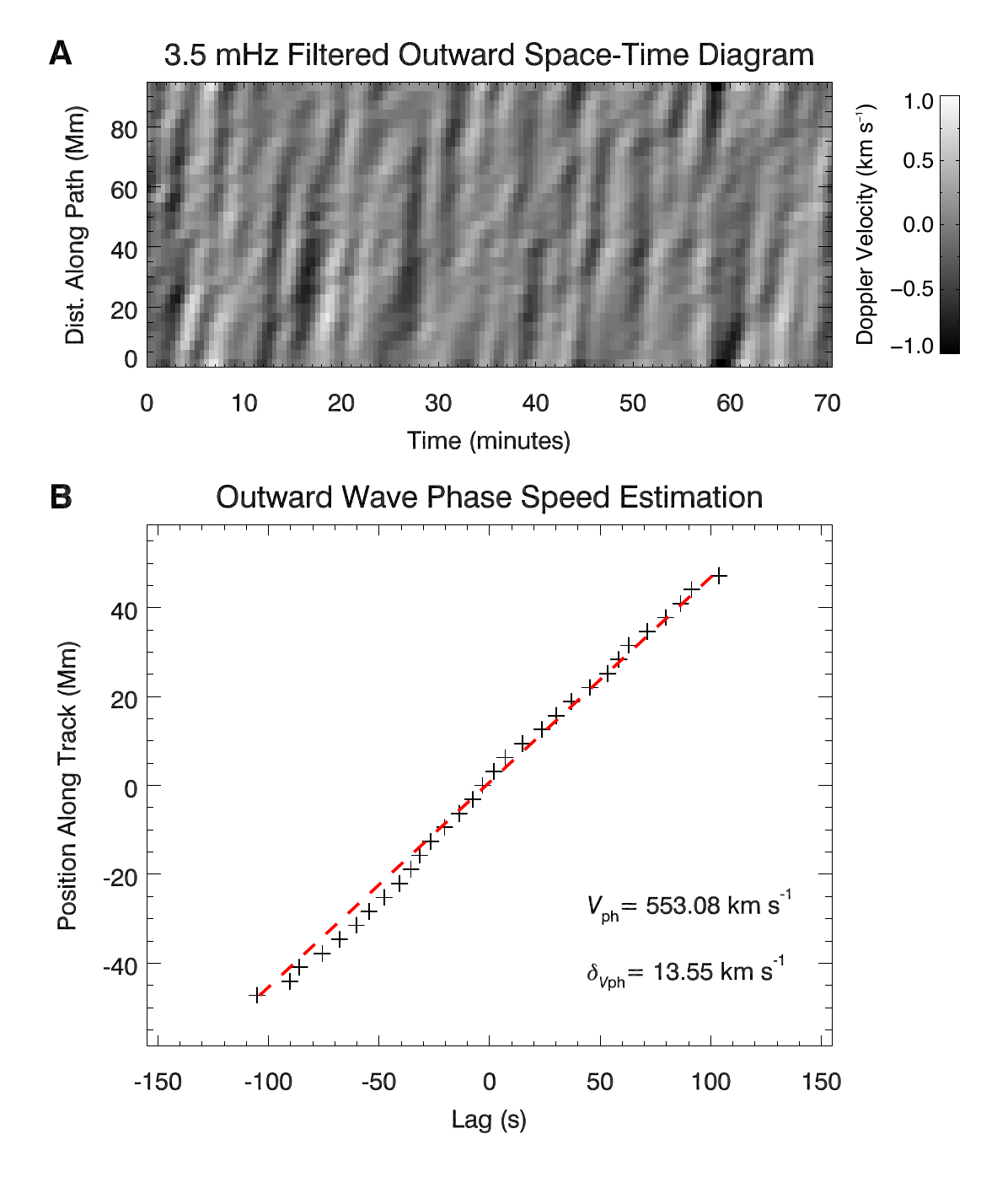}
			\caption{The calculation of phase speed. (A) Space-time diagram for the outward propagating wave along a sample path shown in \cref{fig:fig6}(A). (B) Estimation of the wave phase speed through a linear fitting to the scatter plot of position versus time lag.} 
			\label{fig:fig7}
		\end{figure}

		Based on the derived wave angles, a wave propagation path with a 31-pixel length can be traced for each pixel. The white curve in \cref{fig:fig6}(A) is an example of the traced wave path. A space-time diagram along each wave path was then constructed. A k-$\omega$ diagram was obtained by a fast Fourier transform (FFT) to the space-time diagram. Two branches of the k-$\omega$ diagram correspond to waves that propagate outward and inward. Previous works have shown that the power of the outward propagating wave often dominates over the inward wave, resulting in a less accurate determination of the phase speed of inward wave (e.g., \cite{ref45,ref69}). Therefore, we applied an inverse FFT only to the negative-frequency part of the k-$\omega$ diagram, and obtained a space-time diagram for the outward wave (e.g., \cref{fig:fig7}(A)). A cross correlation of the time series at the center of the path with those at other locations on the path was calculated, resulting in a scatter plot showing the time lag at different locations (e.g., \cref{fig:fig7}(B)). Finally, a linear fitting was applied to the scatter plot to estimate the phase speed and its associated uncertainty. Snapshots of the Doppler velocity image sequences and the wave tracking results for datasets D2 and D3 are presented in \cref{fig:fig2}(F)-(H) and \cref{fig:fig3}(F)-(H), respectively. The phase speeds are mostly in the range of 300 to 700 $\text{km}\ \text{s}^{-1}$, similar to those in dataset D1 \cite{ref52}.
		
		By substituting the derived density and wave phase speed into \cref{Eq:1}, we obtained the magnetic field strength. Since the measured phase speed is the component of the phase speed in the POS, the derived field strength is also the POS component. \cref{fig:fig2}(I)-(J) and \cref{fig:fig3}(I)-(J) show the maps of derived field strengths and associated uncertainties for datasets D2 and D3. The field strengths are mostly 1-5 Gauss and the uncertainties are generally smaller than 15\%, comparable to those in dataset D1 \cite{ref52}. The obtained field strengths are consistent with some estimations of the coronal magnetic field at similar heights using other approaches (e.g., \cite{ref16,ref17,ref61}). As mentioned in our previous work \cite{ref52}, due to the unknown distribution of electron density along the LOS, an additional uncertainty of $\sim$12\% might be present in our derived $B_\text{POS}$. It is also worth mentioning that the current wave tracking method only considers the fitting error when estimating the uncertainty in phase speed. This likely underestimates the uncertainties in phase speed and magnetic field strength at some locations.  
		
\end{multicols}
		
\begin{multicols}{2}

		\subsection{Distributions of coronal magnetic field strength and electron density}
		
\begin{figure}
			\centering
			\includegraphics[scale=0.8]{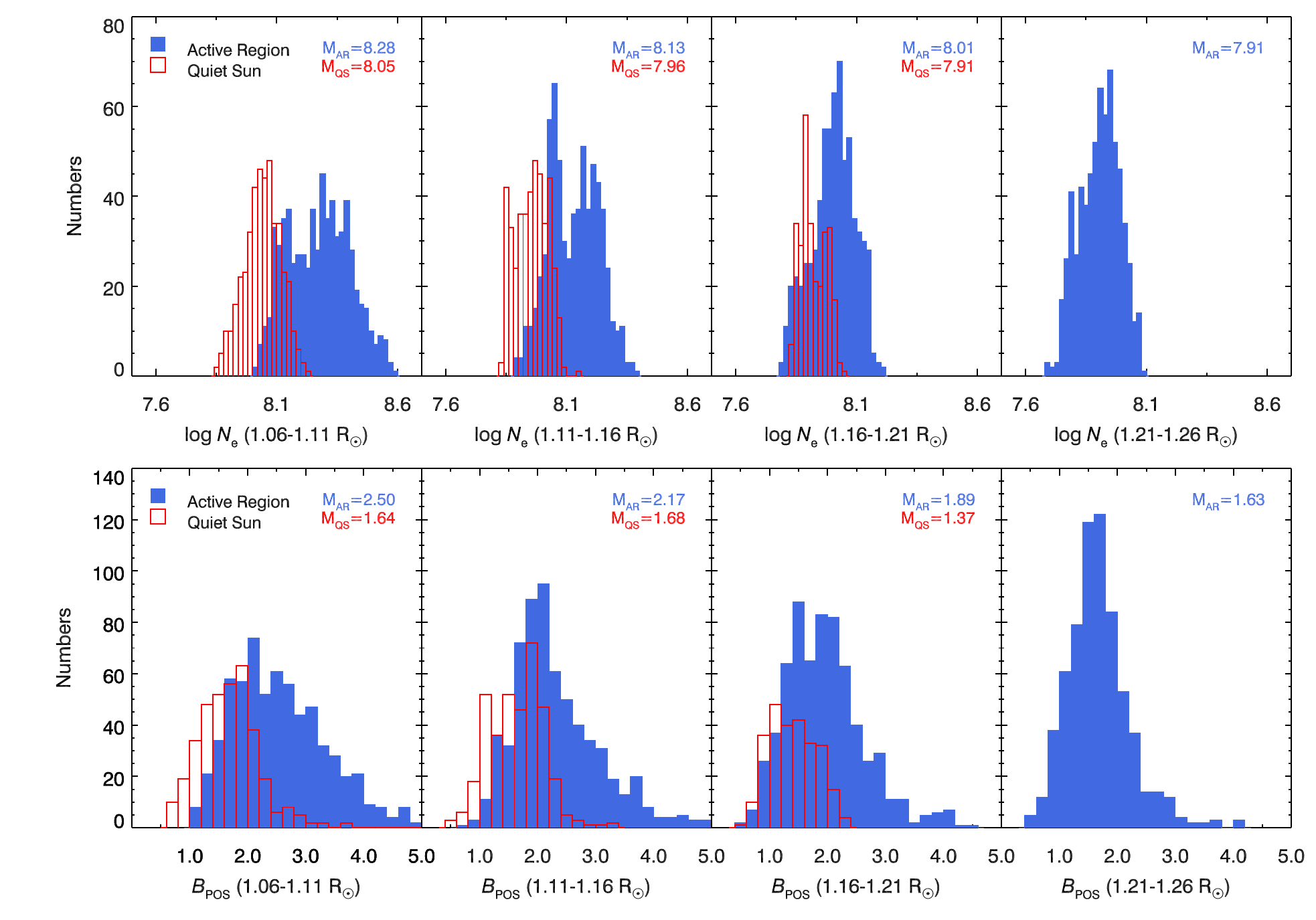}
			\caption{Distributions of the measured electron number density and POS magnetic field strength in different height ranges. The red and blue histograms are for the quiet Sun and active regions, respectively. Median values of parameters for the quiet Sun and active regions are represented by $\text{M}_\text{QS}$ and $\text{M}_\text{AR}$, respectively.} 
			\label{fig:fig8}
		\end{figure}

		Histograms of the measured electron number density ($N_\text{e}$) and magnetic field strength ($B_\text{POS}$) in different height ranges for dataset D1 are presented in \cref{fig:fig8}. The red and blue histograms represent plasma parameters measured from two annulus sectors representing the quiet sun (QS) and another two annulus sectors enclosing active regions (ARs), respectively. The two QS regions are located at the east limb. For the inner boundaries of these two annulus sectors, the ranges of y coordinates are from 394$^{\prime\prime}$ to 506$^{\prime\prime}$ and from -754$^{\prime\prime}$ to -428$^{\prime\prime}$. The two annulus sectors representing ARs are placed at the west limb. For their inner boundaries, the y positions are in the range of 54$^{\prime\prime}$ to 550$^{\prime\prime}$ and -598$^{\prime\prime}$ to -141$^{\prime\prime}$.
		
		It is clear that both the electron density and magnetic field decrease with height, and that both are systematically lower in the QS than in ARs. In addition, the difference between the QS and ARs becomes smaller as the height increases. We have also performed similar analyses for the other two datasets, and found a similar behavior.
		
\begin{figure}
			\centering
			\includegraphics[scale=0.8]{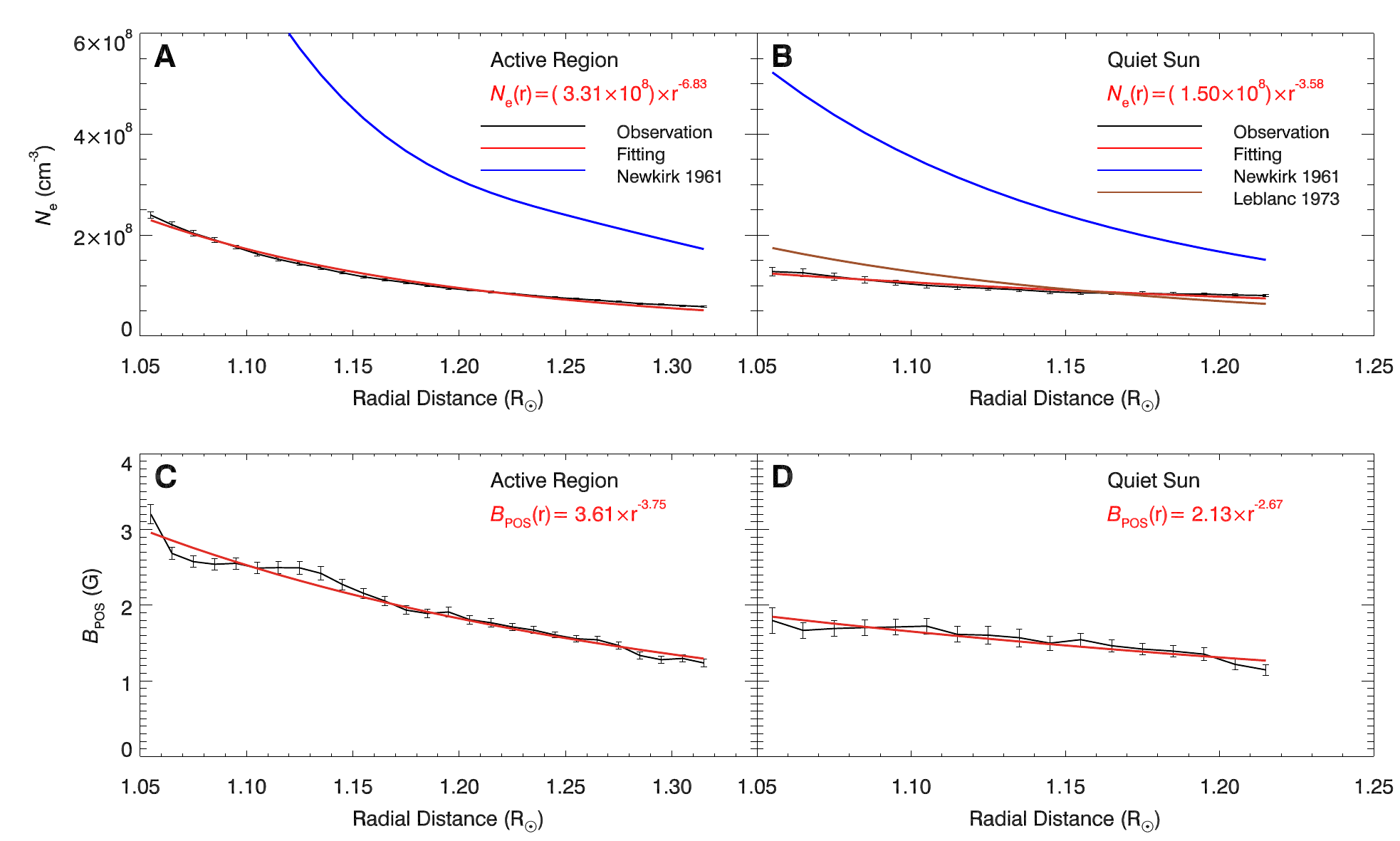}
			\caption{Variations of the measured electron number density (A-B) and POS magnetic field strength (C-D) with height. The black curves and error bars represent measurement results. The red curves show power-law fitting results. The fitting function is also shown in each panel. Results from two density models \cite{ref71, ref72} are also presented for comparison. } 
			\label{fig:fig9}
		\end{figure}

		To show the variations of $N_\text{e}$ and $B_\text{POS}$ with height, we chose the same annulus sectors mentioned above, and averaged the measured physical parameters within each 0.01 $\text{R}_\odot$ height interval in the selected sectors. The height variations of the two parameters and associated standard errors on the mean values are shown in \cref{fig:fig9}. Several previous investigations (e.g., \cite{ref61}) have shown that the variations of the coronal density and magnetic field with height follow a power-law function. Thus, we performed a simple power-law fitting to each measured curve in \cref{fig:fig9}. For both the electron density and magnetic field strength, a larger power-law index has been found in ARs than in the QS, indicating a stronger decrease in ARs. Analyses of the other two datasets have also yielded a similar result. 
		
		For comparison we also show results from two density models \cite{ref71, ref72} in \cref{fig:fig9}(A) and (B). Our measured QS densities appear to be similar to the Leblanc model. The density values from the Newkirk model are several times larger than those in our measurements for both the QS and ARs. However, several recent off-limb observations of the inner corona have provided density values similar to our measurements \cite{ref73, ref74, ref75,ref76}. Also, different active regions could have very different densities, which may partly explain the discrepancy between these measurements and the Newkirk density model.
		
		The measured coronal magnetic field strengths are mostly 1-4 G in the height range of 1.05-1.35 $\text{R}_\odot$. Our derived field strengths are consistent with results from some previous measurements using other techniques. For example, shock observations have revealed field strengths of 1.3-1.5 G in the height range of 1.3-1.5 $\text{R}_\odot$ \cite{ref17} and 1.7-2.1 G in the range of 1.1-1.2 $\text{R}_\odot$ \cite{ref61}. And a spectropolarimetric measurement has yielded a field strength of $\sim$4 G above an active region at the height of 1.1 $\text{R}_\odot$ \cite{ref16}.
\end{multicols}

\begin{multicols}{2}

		\section{Discussion}\label{sec:4}
		Previous measurements of the coronal magnetic field through magnetoseismology normally provided an estimate of only the average field strength or 1D distribution of field strength along an oscillating structure. By applying the technique of magnetoseismology to the pervasive propagating transverse waves observed with CoMP, we now can obtain 2D distributions of the coronal magnetic field (coronal magnetograms), thus marking a leap forward in the application of magnetoseismology. Our analysis demonstrates that both the strength and direction of the coronal magnetic field in the POS can be obtained through actual coronal observations, filling the missing part of the measurements of the Sun's magnetism.
	
		Our technique relies on density diagnostics and wave tracking. For wave tracking, we need to perform a continuous observation of the stronger Fe \sc{xiii}\rm{} 1074.7 nm line for $\sim$1 hour or longer. During this period, normally the Fe \sc{xiii}\rm{} 1079.8 nm line could not be observed with CoMP. Thus, additional time is required to observe both Fe \sc{xiii}\rm{} lines for density diagnostics. In many CoMP observations the Fe \sc{xiii}\rm{} 1079.8 nm line was not used or had a S/N too low to allow a reliable density diagnostic. In such cases, it is still possible to obtain the density using simultaneous observations of the polarized brightness (e.g. \cite{ref33}) with instruments such as K-Cor.
		
		The ubiquitous propagating transverse waves observed with CoMP are identified as kink waves. It has been shown that when the length scale of the waveguide (i.e., radius of the flux tube) is much smaller than the wavelength, the kink wave is nearly incompressible and the restoring force is dominated by magnetic tension \cite{ref47,ref62,ref63}. This indicates that the kink wave is of Alfv\'{e}nic nature at the limit of long wavelength or thin flux tube, which should be the case in our observations. Therefore, these ubiquitous transverse MHD waves are sometimes also called Alfv\'{e}nic waves (e.g., \cite{ref40,ref45,ref47,ref49,ref50,ref77}).
		
		From maps of the wave propagation angle, we see abrupt changes of the angle at several locations. As mentioned in Section 3, some of these abrupt changes (e.g., around y=100$^{\prime\prime}$ in \cref{fig:fig5}(A), y=0$^{\prime\prime}$ in \cref{fig:fig6}(A), y=600$^{\prime\prime}$ in \cref{fig:fig6}(C)) are related to the 180$^{\circ}$ ambiguity, i.e., angles around +90$^{\circ}$ and -90$^{\circ}$ are actually similar angles, thus do not affect the determination of the wave propagation paths and the subsequent calculation of the phase speeds. Some other abrupt changes (e.g., around y=350$^{\prime\prime}$ in \cref{fig:fig6}(C)) are likely related to the complication by possible superimposition of different magnetic structures along the LOS or the low S/N. An anomalous angle may result in an obvious change in the direction of a constructed wave propagation path, meaning that different parts of the constructed path could trace adjacent magnetic field lines. However, from the Doppler velocity image sequence, we can see that the ubiquitous transverse waves often show coherent propagation across at least a few pixels, suggesting that adjacent field lines have similar directions and that waves propagate at similar speeds along adjacent field lines. Considering this, we expect that these anomalous angles may have a limited impact on the calculated phase speeds. We examined several locations of such abrupt changes in wave angles, and found that the extracted wave propagating paths are generally consistent with the loop structures visible from the intensity images, confirming the robustness of our results. A third group of abrupt changes in the maps of wave angle appear near the inner boundary of the FOV (e.g., around y=450$^{\prime\prime}$ in \cref{fig:fig6}(A)). Since there are not enough data points below a near-boundary pixel, the determination of the wave propagation angle at this pixel from the coherence calculation is likely subject to a large uncertainty or even untrustworthy. In addition, there is increased uncertainty near the lower boundary due to guiding errors that cause increased noise just outside the occulting disk. These could result in an obvious deviation of part of the constructed wave path from the actual wave propagation direction. Thus, the calculated wave angles and phase speeds at some locations near the inner boundary should be treated with caution.
		
		Due to some issues with both the wave propagation angle (abrupt changes of wave angle at some locations) and magnetic azimuth (Van Vleck ambiguity and large uncertainty in regions of weak linear polarization), we suggest to use both maps to accurately infer the POS direction of coronal magnetic field. A combination of the two maps allows us to correctly determine the magnetic field directions in regions with less accurate measurements of the wave propagation angle or magnetic azimuth.

		We realize that the current method likely cannot provide magnetograms in coronal holes due to the very low S/N of the two Fe \sc{xiii}\rm{} lines. The Upgraded CoMP (UCoMP, \cite{ref64}) instrument will observe emission from more spectral lines in a larger FOV. Some lines are expected to have emission in the lower-temperature coronal holes, and observations of them might allow us to map the coronal magnetic field in coronal holes. In addition, the higher spatial resolution of UCoMP observations will provide coronal magnetograms at a higher resolution.
		
		With a much smaller FOV and higher spatial resolution, upcoming observations of the local coronal magnetic field from the Daniel K. Inouye Solar Telescope (DKIST, \cite{ref65}) will be complemented by CoMP-like observations of the global or large-scale coronal magnetic field. Moreover, by combining our measurements of the POS field and DKIST measurements of the LOS field through Zeeman effect, information on the coronal vector magnetic field could be obtained.

\end{multicols}

\begin{multicols}{2}
		
		\section{Summary}\label{sec:5}
		
		As a follow-up of our recent application of magnetoseismology in the measurements of the global coronal magnetic field \cite{ref52}, here we have obtained coronal magnetograms from more CoMP observations. Similar to our previous measurements, the electron number density and POS component of magnetic field in the height range of 1.05 $\text{R}_\odot$-1.35 $\text{R}_\odot$ have been found to be mostly $10^{7.5}$-$10^{8.5}$ cm$^{-3}$ and 1-5 Gauss, respectively.
		
		In addition, we have presented maps of the wave propagation direction derived through Stokes-I measurements and the magnetic azimuth derived from linear polarization observations. In general, each of these two maps can reliably reveal the POS distribution of the magnetic field orientation. A combination of these two independent measurements could resolve different uncertainties in the determination of the field direction at some locations.
		
		We have also plotted distributions of the measured electron density and magnetic field strength, and compared their variations with height in the quiet Sun and active regions. A stronger decrease has been found in active regions than in the quiet Sun.
		
		These results demonstrate the great potential of spectroscopic observations with CoMP-like instruments in routine measurements of the coronal magnetic field. Together with simultaneously measured photospheric magnetograms, such coronal magnetograms could provide critical information to advance our understanding of the magnetic coupling between different atmospheric layers as well as the physical mechanisms responsible for many types of dynamics in the solar atmosphere.

		\Acknowledgements{This work is supported by NSFC grants 11825301, 11790304(11790300), Strategic Priority Research Program of CAS (grant XDA17040507), and grant 1916321TS00103201. This material is based upon work supported by the National Center for Atmospheric Research, which is a major facility sponsored by the National Science Foundation under Cooperative Agreement No. 1852977. CoMP is an instrument operated by the National Center for Atmospheric Research. AIA is an instrument on SDO, a mission of NASA’s Living With a Star Program.}

		\bibliographystyle{unsrt}

	\end{multicols}
\end{document}